\documentclass[conference,9pt]{IEEEtran}
\usepackage[T1]{fontenc}
\usepackage{amsmath,amssymb,amsthm}
\usepackage{graphicx,booktabs,array,multirow}
\usepackage{tikz}
\usetikzlibrary{arrows.meta,positioning,fit,calc}
\usepackage{pgfplots}
\pgfplotsset{compat=1.18}
\usepackage{cite}
\usepackage[hidelinks]{hyperref}
\usepackage{xcolor}
\hypersetup{
  pdfauthor={Zhilong Liu, Xiangzhen Zhou, Yunqi Huang, Xin Hong, Pengcheng Zhu, Fanxu Meng, Zhenlong Du},
  pdftitle={CAVQS: Context-Aware Variational Quantum Circuit Synthesis via Unitary Redistribution and Dynamic Reconstruction}
}
\newcommand{\CAVQS}{CAVQS}
\newcommand{\cx}{N_{\mathrm{CX}}}
\newcommand{\cost}{\overline N_{\mathrm{CX}}}
\newcommand{\pre}{U_{\mathrm{pre}}}
\newcommand{\dyn}{\widehat U_{\mathrm{dyn}}}
\newcommand{\supp}{\operatorname{supp}}
\newcommand{\tr}{\operatorname{Tr}}
\newtheorem{proposition}{Proposition}

\newlength{\resultintegerwidth}
\newcommand{\resultentry}[3]{%
  \makebox[\resultintegerwidth][r]{#1}.#2#3%
}

\title{CAVQS: Context-Aware Variational Quantum Circuit Synthesis via Unitary Redistribution and Dynamic Reconstruction}

\author{
\IEEEauthorblockN{
Zhilong Liu\textsuperscript{1},
Xiangzhen Zhou\textsuperscript{1,2,*},
Yunqi Huang\textsuperscript{3},
Xin Hong\textsuperscript{4},\\
Pengcheng Zhu\textsuperscript{5},
Fanxu Meng\textsuperscript{1},
Zhenlong Du\textsuperscript{1}
}
\IEEEauthorblockA{
\textsuperscript{1}Nanjing Tech University,
Nanjing, China\\
\textsuperscript{2}Arclight Quantum Computing Inc.,Beijing, China\\
\textsuperscript{3}Quantum Science Center of
Guangdong-Hong Kong-Macao Greater Bay Area,
Shenzhen, China\\
\textsuperscript{4}Institute of Software,
Chinese Academy of Sciences,
Beijing, China\\
\textsuperscript{5}Taizhou University,
Taizhou, China\\
\textsuperscript{*}Corresponding author:
Xiangzhen Zhou
(\texttt{xiangzhenzhou@njtech.edu.cn})
}
}

\begin{document}
\maketitle
\begin{abstract}
Variational quantum circuit synthesis jointly searches circuit structure and continuous parameters to reduce costly two-qubit gates, but block-based workflows typically instantiate blocks independently, limiting simplification across block boundaries. Moreover, fully unitary synthesis does not exploit opportunities created by permitted output measurements and classical feedforward. We present \CAVQS{}, a context-aware variational synthesis framework that combines \emph{unitary redistribution} and \emph{dynamic reconstruction}. First, CAVQS uses QR decomposition to transfer unitary factors between overlapping blocks, preserving the overall transformation while reshaping local synthesis targets to enable cross-block reductions. Second, for blocks that precede permitted terminal measurements, CAVQS constructs ancilla-free dynamic implementations with a CNOT-budgeted pre-measurement search and analytic branch-unitary construction, thereby minimizing expected CNOT cost. Across random-circuit targets, CAVQS reduces mean two-qubit gate counts by 28.79\%--42.52\%, 2.05\%--52.62\%, and 30.51\%--74.64\% relative to QFAST, SQUANDER, and Fast-USYN, respectively, on common successful instances. On 4--6-qubit quantum algorithm benchmarks, the corresponding reductions are 27.27\%--36.77\%, 20.65\%--69.10\%, and 36.31\%--71.88\%, respectively.
\end{abstract}

\section{Introduction}
Quantum circuit synthesis maps a desired unitary operation to a series of implementable gates. The final number of gates influences runtime resources, so synthesized circuits are particularly important on noisy hardware, where two-qubit gates introduce significant errors and high execution overhead~\cite{preskill}. Analytical decompositions provide general constructions for arbitrary unitaries~\cite{shende,krol,mottonen,bullock}, while numerical synthesis searches for implementations with fewer gates.

Variational synthesis combines a search over circuit structures with continuous optimization of gate parameters to approximate a target unitary within a prescribed error threshold. Its main challenge is to find gate-efficient circuits with manageable search cost. QSearch searches over elementary-gate structures~\cite{qsearch}, and LEAP scales numerical synthesis through incremental search~\cite{leap}. SQUANDER and Fast-USYN improve gate-count optimization and candidate selection~\cite{squander,fastusyn}, with Fast-USYN drawing on QuCT's models of circuit context~\cite{quct}. An SRBB-based approach uses a structured operator basis for approximate unitary synthesis~\cite{srbb}. Related circuit-optimization methods include hierarchical resynthesis~\cite{qgo}, parameterized instantiation~\cite{paraminst,qfactor}, superoptimization~\cite{quartz}, combined rewriting and resynthesis~\cite{guoq}, circuit resizing~\cite{circuitresizing}, and fidelity-oriented approximation~\cite{quest}. Block-based methods like QFAST instead optimize sequences of smaller unitary blocks before instantiating them as elementary gates~\cite{qfast}.

This block-based workflow offers two opportunities to reduce gate counts by using circuit context. The first opportunity arises at block boundaries. When blocks are instantiated separately, each synthesizer optimizes a fixed local target, limiting simplification across neighboring blocks. Yet overlapping blocks can exchange a unitary factor on their shared qubits without changing the combined transformation. Such transfers can improve subsequent gate-level instantiation and reduce the total gate count.

The second opportunity arises when output measurements are permitted. A purely unitary implementation keeps operations coherent even when a qubit's measured value could control later operations classically. After that qubit's last required coherent interaction, a permitted terminal measurement can be moved earlier, creating conditional branches on the remaining qubits. Such a dynamic implementation changes the cost objective: gates before measurement execute on every run, whereas branch gates contribute according to their outcome probabilities. Dynamic synthesis must therefore choose both a low-cost pre-measurement circuit and suitable branch operations. Prior work uses dynamic circuits for the quantum Fourier transform~\cite{dynamicqft} and qubit reuse~\cite{qubitreuse}, and explores reinforcement learning for dynamic unitary synthesis~\cite{valcarce2025}. AC/DC uses numerical synthesis to construct dynamic circuits with measurement and classical feedforward, including an ancilla-assisted protocol for unitary implementation~\cite{acdc}. It emphasizes reducing circuit depth, but its reported implementation restricts each post-measurement branch to a single layer of single-qubit gates, limiting the operations that can be moved into these branches.

We present \CAVQS{}, which, to the best of our knowledge, is the first ancilla-free variational synthesis framework to exploit permitted terminal measurements in the synthesis of arbitrary unitaries. It addresses these opportunities through two core steps. \emph{Unitary redistribution} changes the assignment of operations to overlapping blocks before gate instantiation. \emph{Dynamic reconstruction} then replaces eligible blocks with pre-measurement circuits and conditional branches to reduce expected execution cost. CAVQS builds on QFAST for block generation and QSearch for gate instantiation. Our contributions are:

\begin{itemize}
\item \textbf{Unitary redistribution.}
We propose a modification to the initial block placement and a method for redistributing unitary factors between overlapping blocks. {We use QR decomposition \cite{de2020quantum} to select the factors to transfer.} These transfers preserve the overall transformation while changing the local synthesis targets, allowing simplifications to occur across block boundaries.

\item \textbf{Dynamic reconstruction.}
We design a CNOT-budgeted pre-measurement search for blocks selected according to terminal measurement positions, allowing entangling gates to be shifted into conditionally executed branches. We adapt QFactor's singular value decomposition (SVD)-based optimization of local unitary gates in static circuits~\cite{qfactor} to analytically construct locally optimal conditional branch unitaries for a fixed pre-measurement circuit and measurement position. We then further refine these branches via recursive reconstruction and by absorbing them into later blocks, accepting substitutions that lower the expected cost while still satisfying the full-circuit error bound.

\item \textbf{Experimental evaluation.}
We evaluate CAVQS against QFAST, QFAST-SM, SQUANDER, and Fast-USYN.
On benchmarks with unrestricted connectivity, per-size comparisons on common successful targets show that CAVQS achieves lower mean two-qubit gate counts than the baselines, with reductions of up to 74.64\% (relative to Fast-USYN on five-qubit targets) on random-circuit targets and 71.88\% (relative to Fast-USYN on five-qubit benchmarks) on quantum algorithm benchmarks.
Ablation studies show that combining unitary redistribution and dynamic reconstruction reduces the mean two-qubit counts by 17.75\% and 21.80\% over static-only and dynamic-only variants, respectively. With line, ring and star connectivity constraints, CAVQS further reduces mean two-qubit counts by 40.48\%, 34.87\%, and 13.50\% relative to QFAST.

\end{itemize}

\section{Problem Formulation}\label{sec:problem}
\subsection{Target and Measurement Permissions}
Given an $n$-qubit target unitary $U_T\in U(2^n)$, a set $Q_M$ of output qubits that may be measured in the computational basis, and an error threshold $\epsilon$, we seek a circuit built from parameterized single-qubit $U3$ gates and CNOT gates. Here, $U(2^n)$ is the group of $2^n\times2^n$ unitary matrices. When $Q_M=\varnothing$, the circuit is static and approximates $U_T$. When output measurements are permitted, the circuit may also use measurements and classical feedforward to form a dynamic implementation that approximates the measurement probabilities and conditional outputs of applying $U_T$ and then measuring the designated outputs.

To evaluate a dynamic circuit $C$, we use its deferred-measurement representation $\widehat U_C$, which replaces classical branch selection with coherent control~\cite{deferred}. The circuit executes with physical measurements, but this representation allows us to evaluate it as a unitary matrix. We require the global-phase-insensitive loss to satisfy
\begin{equation}
 \Delta(\widehat U_C,U_T)=1-\frac{|\tr(U_T^\dagger\widehat U_C)|}{2^n},
 \qquad \Delta(\widehat U_C,U_T)\leq\epsilon.
 \label{eq:loss}
\end{equation}
Here, $\tr$ denotes the matrix trace, $\dagger$ denotes the conjugate transpose, and smaller $\Delta$ indicates a closer approximation. This full-matrix test is sufficient for the intended measurement behavior and also distinguishes representations that differ only in branch phases, even when their measured outputs agree.

\subsection{Optimization Objective}
We minimize the expected CNOT count
\begin{equation}
 \cost(C)=\sum_h p(h)\cx(C_h),\label{eq:cost}
\end{equation}
where $h$ denotes a complete measurement history, $p(h)$ is its probability, $C_h$ is the executed path, and $\cx$ counts its CNOT gates. For a static circuit, Eq.~\eqref{eq:cost} reduces to the CNOT count. A simulator estimates the probabilities for a fixed reference input, specified in the experimental setup, while Eq.~\eqref{eq:loss} checks synthesis accuracy over the full unitary.

\section{Motivating Example and Framework Overview}\label{sec:overview}
\subsection{Two Opportunities in One Block Sequence}
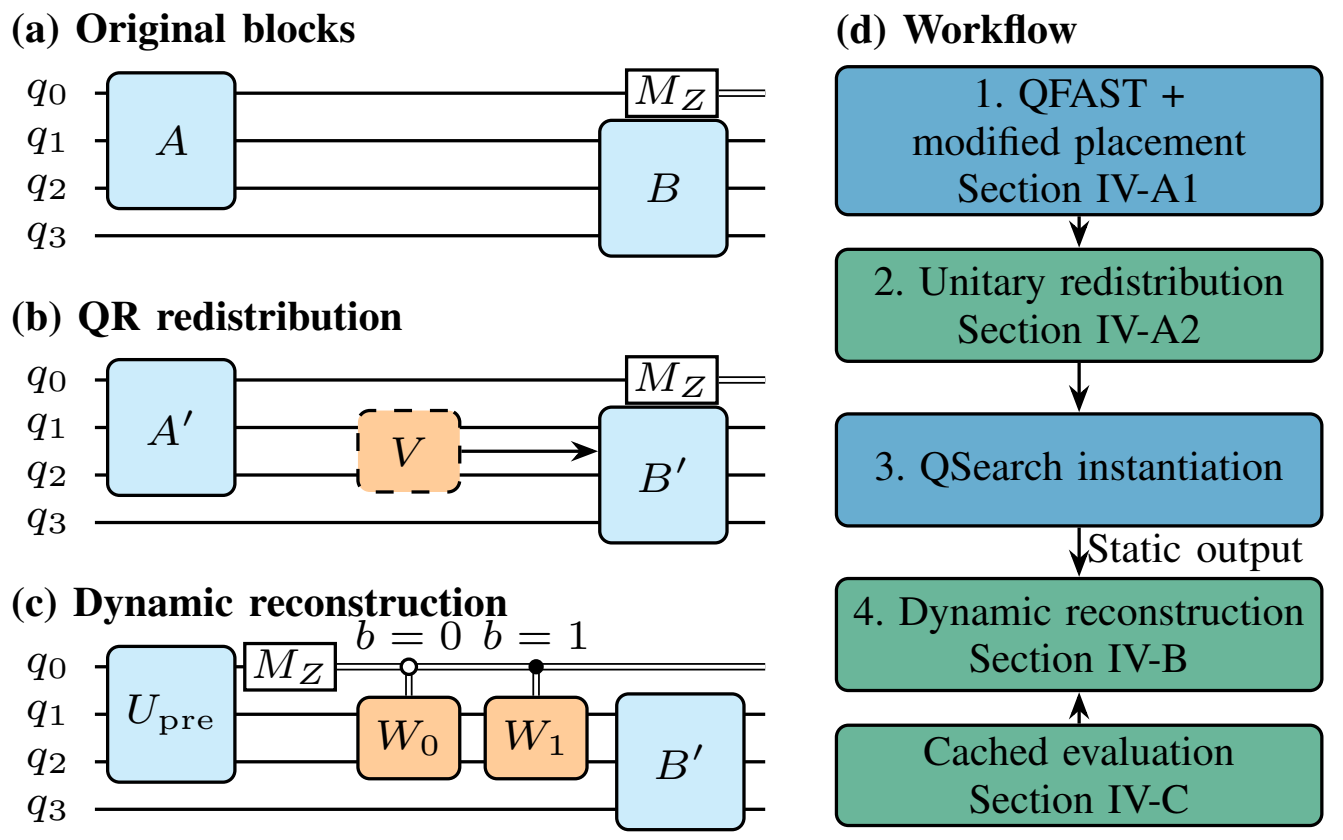
\begin{figure}[t]
    \centering
    \resizebox{0.98\columnwidth}{!}{%
    \begin{tikzpicture}[
        x=1cm,y=1cm,font=\scriptsize,
        quantum wire/.style={line width=0.55pt},
        classical wire/.style={line width=0.3pt,double,double distance=0.65pt},
        quantum block/.style={draw,line width=0.55pt,rounded corners=2pt,
            fill=cyan!18,minimum width=0.75cm,minimum height=0.80cm,inner sep=1pt},
        branch block/.style={quantum block,fill=orange!40,
            minimum width=0.60cm,minimum height=0.48cm},
        measurement/.style={draw,fill=white,line width=0.55pt,
            minimum width=0.50cm,minimum height=0.24cm,inner sep=1pt},
        flow arrow/.style={-{Stealth[length=1.6mm,width=1.2mm]},line width=0.55pt},
        stage/.style={draw,line width=0.55pt,rounded corners=2pt,
            text width=2.72cm,minimum height=0.66cm,align=center,inner sep=2pt},
        panel label/.style={font=\scriptsize\bfseries,align=left,
            anchor=west,inner sep=0pt}
    ]
        \begin{scope}
            \node[panel label] at (0,0.36) {(a) Original blocks};
            \foreach \q in {0,...,3} {
                \draw[quantum wire] (0.50,{-0.28*\q}) -- (4.45,{-0.28*\q});
                \node[anchor=east,inner sep=1pt] at (0.40,{-0.28*\q}) {$q_{\q}$};
            }
            \node[quantum block] at (0.95,-0.28) {$A$};
            \node[quantum block] at (3.85,-0.56) {$B$};
            \node[measurement] (measure-a) at (3.9,0) {$M_Z$};
            \draw[classical wire] (measure-a.east) -- (4.45,0);
        \end{scope}

        \begin{scope}[yshift=-1.69cm]
            \node[panel label] at (0,0.36) {(b) QR redistribution};
            \foreach \q in {0,...,3}
                \node[anchor=east,inner sep=1pt] at (0.40,{-0.28*\q}) {$q_{\q}$};
            \foreach \q in {0,...,3}
                \draw[quantum wire] (0.50,{-0.28*\q}) -- (4.45,{-0.28*\q});
            \node[quantum block] at (0.95,-0.28) {$A'$};
            \node[branch block,dashed] (transfer-v) at (2.35,-0.42) {$V$};
            \node[quantum block] (redistributed-b) at (3.85,-0.56) {$B'$};
            \draw[flow arrow] (transfer-v.east) -- (redistributed-b.west |- transfer-v.east);
            \node[measurement] (measure-b) at (3.9,0) {$M_Z$};
            \draw[classical wire] (measure-b.east) -- (4.45,0);
        \end{scope}

        \begin{scope}[yshift=-3.38cm]
            \node[panel label] at (0,0.36) {(c) Dynamic reconstruction};
            \foreach \q in {0,...,3}
                \node[anchor=east,inner sep=1pt] at (0.40,{-0.28*\q}) {$q_{\q}$};
            \foreach \q in {1,...,3}
                \draw[quantum wire] (0.50,{-0.28*\q}) -- (4.45,{-0.28*\q});
            \draw[quantum wire] (0.50,0) -- (1.65,0);
            \node[quantum block] at (0.95,-0.28) {$U_{\mathrm{pre}}$};
            \node[measurement] (measure-c) at (1.65,0) {$M_Z$};
            \draw[classical wire] (measure-c.east) -- (4.45,0);
            \node[branch block] (branch-zero) at (2.35,-0.42) {$W_0$};
            \node[branch block] (branch-one) at (3.10,-0.42) {$W_1$};
            \draw[classical wire] (2.35,0) -- (branch-zero.north);
            \draw[classical wire] (3.10,0) -- (branch-one.north);
            \draw[fill=white,line width=0.55pt] (2.35,0) circle[radius=0.045cm];
            \fill (3.10,0) circle[radius=0.045cm];
            \node[anchor=south,inner sep=1pt] at (2.35,0.06) {$b=0$};
            \node[anchor=south,inner sep=1pt] at (3.10,0.06) {$b=1$};
            \node[quantum block] at (3.95,-0.56) {$B'$};
        \end{scope}

        \node[panel label] at (4.86,0.36) {(d) Workflow};
        \definecolor{CAVQSBackendFill}{HTML}{68ADD0}
        \definecolor{CAVQSMethodFill}{HTML}{6AB69A}
        \node[stage,fill=CAVQSBackendFill] (stage-one) at (6.30,-0.28)
            {1. QFAST + modified placement\\Section~\ref{sec:layout}};
        \node[stage,fill=CAVQSMethodFill] (stage-two) at (6.30,-1.25)
            {2. Unitary redistribution\\Section~\ref{sec:absorption}};
        \node[stage,fill=CAVQSBackendFill] (stage-three) at (6.30,-2.22)
            {3. QSearch instantiation};
        \node[stage,fill=CAVQSMethodFill] (stage-four) at (6.30,-3.19)
            {4. Dynamic reconstruction\\Section~\ref{sec:dynamic}};
        \draw[flow arrow] (stage-one.south) -- (stage-two.north);
        \draw[flow arrow] (stage-two.south) -- (stage-three.north);
        \draw[flow arrow] (stage-three.south) -- node[midway,right,inner sep=1pt] {Static output} (stage-four.north);
        \node[stage,fill=CAVQSMethodFill,minimum height=0.36cm]
            (cached-evaluation) at (6.30,-4.02)
            {Cached evaluation\\Section~\ref{sec:implementation}};
        \draw[flow arrow] (cached-evaluation.north) -- (stage-four.south);
    \end{tikzpicture}%
    }
    \caption{A four-qubit example and the CAVQS workflow.
    (a) Original blocks. (b) QR redistribution transfers the dashed factor $V$ from $A$ into $B$.
    (c) Computational-basis measurement $M_Z$ selects exactly one branch $W_b$ on $q_1,q_2$.
    Double lines carry classical outcomes.
    (d) Synthesis workflow: blue denotes off-the-shelf synthesis backends with modified block placement in Stage 1 and green denotes CAVQS implementation.}
    \label{fig:framework}
\end{figure}

Figure~\ref{fig:framework}(a) shows four qubits labeled $q_0,\ldots,q_3$ and consecutive three-qubit blocks $A(q_0,q_1,q_2)$ and $B(q_1,q_2,q_3)$, followed by a permitted computational-basis measurement of $q_0$. A unitary $V\in U(4)$ on their shared qubits can be reassigned by setting
\begin{equation}
 A'=(I_{q_0}\otimes V^\dagger)A,\qquad
 B'=B(V\otimes I_{q_3}).\label{eq:transfer}
\end{equation}
Here, $I_{q_i}$ is the identity on qubit $q_i$, and tildes denote embeddings into the full four-qubit space, where $\widetilde B'\widetilde A'=\widetilde B\widetilde A$. The reassigned blocks in Fig.~\ref{fig:framework}(b) preserve the complete transformation and the terminal measurement while changing the two matrices passed to the gate synthesizer. The transferred factor can include entangling operations on the shared qubits.

Because $B'$ does not act on $q_0$, we can reconstruct $A'$ as shown in Fig.~\ref{fig:framework}(c): apply a pre-measurement unitary $\pre$, measure $q_0$, and apply $W_0$ or $W_1$ to $q_1,q_2$ for outcome $b=0$ or $b=1$, respectively. Only the branch selected by $b$ executes before $B'$. With $B'$ unchanged, the expected cost is
\begin{equation}
 \cx(\pre)+\sum_{b=0}^1p_b\cx(W_b)+\cx(B'),\label{eq:examplecost}
\end{equation}
where $p_b$ is the probability of measurement outcome $b$, and $\cx$ applied to a unitary counts the CNOT gates in its chosen implementation. Furthermore, the branch unitaries $W_0$ and $W_1$ can be separately absorbed into $B'$, yielding outcome-dependent downstream blocks for further optimization.

\subsection{Framework Overview}

The four stages in Fig.~\ref{fig:framework} combine existing synthesis backends with CAVQS's redistribution and reconstruction procedures.

\emph{Stage 1: block generation.} \CAVQS{} uses QFAST's block representation and optimizer to synthesize $U_T^\dagger$, with initial block placements chosen to maximize overlap between adjacent blocks.
 It decomposes the result into blocks of $s$ qubits, where $s$ is the chosen block width, and inverts the sequence. Section~\ref{sec:layout} describes the placement policy.

\emph{Stage 2: redistribution.} \CAVQS{} merges compatible blocks or uses QR decomposition to transfer a factor on shared qubits to a later block. Section~\ref{sec:absorption} gives the selection and transfer rules.

\emph{Stage 3: instantiation.} QSearch converts the processed blocks to elementary gates. The assembled static circuit is checked for accuracy and provides the gate-count baseline for reconstruction.

\emph{Stage 4: dynamic reconstruction.} \CAVQS{} uses QSearch to find pre-measurement circuits and SVD to construct their branches, then recursively reconstructs eligible branches (Section~\ref{sec:dynamic}). A replacement is accepted when it reduces expected cost and the complete circuit meets the error threshold. Caching reduces repeated evaluation work during this search (Section~\ref{sec:implementation}).

\section{CAVQS Method}
This section describes the placement and redistribution policies in Stages~1--2, dynamic reconstruction in Stage~4, and the optimizations used to evaluate candidates.

\subsection{Block Generation and Unitary Redistribution}
\subsubsection{Block generation}\label{sec:layout}
We improve QFAST's block placement to support the subsequent unitary redistribution and dynamic reconstruction. Specifically, we arrange the initial $s$-qubit blocks so that neighboring blocks share $s-1$ qubits. We use this layout to synthesize $U_T^\dagger$, while freely optimizing the block parameters. If the initial sequence is insufficient, the search adds blocks at unrestricted locations. We then recursively decompose the result to width $s$, reverse the block order, and take the conjugate transpose of each block. This gives a sequence $(U_1,\ldots,U_N)$ approximating $U_T$, where $N$ is the number of blocks and $U_k$ is the $k$th block in execution order. This layout provides more opportunities for optimization in both unitary redistribution and dynamic reconstruction.

\subsubsection{QR-based unitary redistribution}\label{sec:absorption}
We use QR decomposition to choose the unitary factor transferred between overlapping blocks. In the example in Fig.~\ref{fig:framework}, we first select the four rows of $A$ whose nonshared output is $q_0=0$. These rows are linearly independent because $A$ is unitary. We then select four columns one at a time. At each step, we orthogonalize each remaining column against the previously selected columns and choose the column with the largest resulting norm. Ties are resolved by the original column index. Finally, we apply QR decomposition to the resulting full-rank matrix $M\in\mathbb C^{4\times4}$ formed from the selected original columns:
\begin{equation}
 M=VR,\label{eq:qr}
\end{equation}
where $V\in U(4)$ is unitary and $R$ is upper triangular. We require the diagonal entries of $R$ to be positive and real so that $V$ is unique.
Applying Eq.~\eqref{eq:transfer} then makes the selected submatrix of $A'$ equal to $R$ and transfers $V$ to $B$, yielding $B'$.

Both updated blocks remain unitary. Their combined operation is unchanged because the inserted factors cancel:
\begin{equation*}
\widetilde B'\widetilde A'
=\widetilde B(I_{q_0}\otimes VV^\dagger\otimes I_{q_3})\widetilde A
=\widetilde B\widetilde A.
\end{equation*}
More generally, for $r$ shared qubits, we select $2^r$ rows and $2^r$ independent columns and transfer an $r$-qubit unitary. Thus, neighboring $s$-qubit blocks that share $s-1$ qubits can exchange an $(s-1)$-qubit unitary. This factor can include entangling operations when $s-1\geq2$. We use QR as a heuristic because it introduces zeros into the selected submatrix, consistent with viewing synthesis as reducing a unitary toward the identity.

\smallskip\noindent\emph{Numerical illustration.}
Consider a three-qubit unitary $A$ whose four rows with $q_0=0$ form
\begingroup\setlength{\arraycolsep}{2pt}
\begin{equation*}
X=\frac{1}{25\sqrt{2}}\begin{bmatrix}
9&-12&20&0&20&0&15&0\\
16&12&0&15&0&15&0&20\\
16&12&0&-15&0&15&0&-20\\
9&-12&-20&0&20&0&-15&0
\end{bmatrix}.
\end{equation*}
\endgroup
Using zero-based column indices, columns $2$, $4$, and $7$ have equal norms of $0.8$ and are mutually orthogonal, so they are selected first in index order. After orthogonalizing the remaining columns against these three, column $0$ has the largest resulting norm of $0.64$. Thus, $J=(2,4,7,0)$. The resulting matrix for QR decomposition is
\begingroup\setlength{\arraycolsep}{2pt}
\begin{equation*}
M=X_{:,J}=\frac{1}{25\sqrt{2}}
\begin{bmatrix}
20&20&0&9\\
0&0&20&16\\
0&0&-20&16\\
-20&20&0&9
\end{bmatrix}=VR.
\end{equation*}
\endgroup
Thus, the selected submatrix of $A'$ becomes
$V^\dagger M=R$, which is upper triangular.

We scan the blocks in execution order. We first try to merge blocks that act on the same qubits, whenever the merge is legal. If no merge is possible, we try a QR transfer to a later block that shares some of its qubits. When several target blocks are available, we prefer those sharing more qubits. For nonadjacent blocks, the transferred factor must commute with every intervening block. We continue the scan using the updated blocks, then use QSearch to synthesize them into the static circuit $C_{\mathrm{base}}$. We keep the block matrices as targets for dynamic reconstruction and use their CNOT counts in this static circuit to bound the search budgets.

\subsection{Dynamic Reconstruction}\label{sec:dynamic}
For each eligible block, we first search for a pre-measurement circuit and then construct and synthesize its branches. We accept the replacement only if it reduces the expected cost and the full circuit meets the error threshold.
\subsubsection{Measurement eligibility}
To select measurement qubits for the dynamic reconstruction of each block, we identify qubits that belong to the block, are permitted to be measured at the output, and are not used by any later block. To find them, we scan the blocks backward from $U_N$, keeping the qubits used by later blocks in a set $Q_{\mathrm{used}}$, initially empty. For block $U_k$, the eligible measurement qubits are
\begin{equation}
 M_k=(Q_M\cap\supp(U_k))\setminus Q_{\mathrm{used}},\label{eq:eligible}
\end{equation}
where $\supp(U_k)$ is the set of qubits on which $U_k$ acts. After recording $M_k$, we add $\supp(U_k)$ to $Q_{\mathrm{used}}$ and continue the backward scan. We then process the recorded blocks in execution order. In Fig.~\ref{fig:framework}, $q_0$ is eligible in $A'$, whereas $B'$ still needs $q_1,q_2$.

\subsubsection{Budgeted pre-measurement search}
At each reconstruction node, CNOTs before measurement execute whenever the node is reached, whereas branch CNOTs execute only when the corresponding branch is selected. To reduce this cost, we limit the pre-measurement CNOT count with a budget $K$. The upper bound is $\cx(C_k)$, the CNOT count of block $U_k$'s implementation $C_k$ in the static circuit $C_{\mathrm{base}}$.

For each eligible measurement qubit $q$, we place it first in the tensor product ordering and write the candidate's deferred-measurement representation $\dyn$ as
\begin{equation}
 \dyn=\left(\sum_{b=0}^1|b\rangle\langle b|\otimes W_b\right)\pre,
 \label{eq:dynamic}
\end{equation}
where $\pre$ acts before measurement and $W_b$ is the branch unitary for outcome $b$. The expected cost is therefore $\cx(\pre)+\sum_b p_b\cx(W_b)$, with $\cx(\pre)\leq K$.

We start with a CNOT budget of $K=\lfloor\cx(C_k)/2\rfloor$. For each budget, we first set $W_0=W_1=I$, where $I$ is the identity on the unmeasured qubits, and use QSearch to search gate structures and optimize the $U3$ parameters of $\pre$. QSearch evaluates each candidate in the full circuit and keeps the one with the lowest loss within the budget. We then fix $\pre$ and compute its branch unitaries analytically, as described next. After evaluating the completed candidate at the initial budget, we increase $K$ by one if the full-circuit error exceeds $\epsilon$, stopping when the threshold is met or the static CNOT count is reached. Otherwise, we decrease $K$ by one until failure, retaining the last successful candidate. The candidate is accepted only if it also reduces the expected CNOT count.

\subsubsection{SVD-based branch construction}
For a fixed measurement location and $\pre$, we construct branches for the target block $U_k$ from
\begin{equation}
 P_b=\langle b|\otimes I,\qquad R_b=P_bU_k\pre^\dagger P_b^\dagger.
 \label{eq:branchmatrix}
\end{equation}
Here, $P_b$ selects outcome $b$, so $R_b$ is the corresponding diagonal block of $U_k\pre^\dagger$. We compute the SVD $R_b=A_b\Sigma_bB_b^\dagger$, where $A_b,B_b$ are unitary and $\Sigma_b$ contains the singular values. We then set
\begin{equation}
 W_b=A_bB_b^\dagger.\label{eq:branchsvd}
\end{equation}

\begin{proposition}
For a fixed $\pre$ and measurement location, Eq.~\eqref{eq:branchsvd} minimizes the local loss $\Delta(\dyn,U_k)$ over all choices of branch unitaries on the remaining qubits.
\end{proposition}
\begin{proof}
Using cyclic invariance of the trace in Eq.~\eqref{eq:dynamic}, we bound the trace overlap by
\begin{align}
 |\tr(U_k^\dagger\dyn)|
 &=\left|\sum_b\tr(W_b^\dagger R_b)\right|\nonumber\\
 &\leq\sum_b|\tr(W_b^\dagger R_b)|
 \leq\sum_b\tr(\Sigma_b).\label{eq:bound}
\end{align}
Equation~\eqref{eq:branchsvd} makes each summand equal to $\tr(\Sigma_b)\geq0$ and attains the bound.
\end{proof}
The minimum local loss is $1-2^{-s}\sum_b\|R_b\|_*$, where the nuclear norm $\|R_b\|_*$ is the sum of singular values. For an $s$-qubit block, the two $2^{s-1}\times2^{s-1}$ matrices directly determine the $(s-1)$-qubit branch unitaries. However, this local optimum does not guarantee that the full circuit meets the error threshold, so we still need to check Eq.~\eqref{eq:loss}.

\subsubsection{Branch implementation and recursive reconstruction}

If the full-circuit loss after replacing the block is at most $\epsilon$, we implement each single-qubit branch as a single-qubit gate and use QSearch to synthesize branches with two or more qubits.\footnote{Since QSearch uses a global-phase-insensitive loss, independently instantiated branches may introduce relative phase offsets in the deferred-measurement representation. CAVQS compensates for these offsets by applying an $R_z$ rotation to the measured qubit before measurement, with the angle determined from trace overlaps between the SVD-derived and instantiated branch unitaries.}

Each branch is itself a smaller quantum circuit, so it can also be reconstructed using measurement and conditional operations to further reduce the expected CNOT count. We repeat this procedure when the branch has at least two qubits and contains a qubit that is permitted to be measured and is no longer needed by later operations, either inside or outside the branch. For example, in Fig.~\ref{fig:framework}(c), $B'$ still performs quantum operations on $q_1,q_2$, so we must leave both qubits unmeasured in $W_0$ and $W_1$. Each recursive step measures one qubit, leaving one fewer qubit in the new branches. We accept each replacement only if the full-circuit loss remains at most $\epsilon$ and the expected CNOT count decreases.

The simulator computes the probability of each complete measurement history. For a reconstruction node $D$ with incoming conditional state $\rho$, let $\mathcal C(D\mid\rho)$ denote its expected CNOT count, computed recursively as
\begin{equation}
 \mathcal C(D\mid\rho)=\cx(\pre)+\sum_b p_b\mathcal C(D_b\mid\rho_b),
 \label{eq:recursivecost}
\end{equation}
where $p_b$ is the probability of outcome $b$ at node $D$, $D_b$ is the corresponding branch, and $\rho_b$ is the input state to $D_b$ conditioned on that measurement outcome.

\subsubsection{Circuit postprocessing and candidate selection}
Operations moved from the end of the pre-measurement circuit into the branches can be combined with each branch's existing operations, creating opportunities to reduce the expected CNOT count. To find such operations, we inspect the gate-level circuit for $\pre$ produced by QSearch. If its final operation $T$ acts only on the unmeasured qubits, the circuit has the form $\pre=(I\otimes T)\pre'$, where $\pre'$ represents the remaining gates. We then remove $T$ from $\pre$ and update each branch from $W_b$ to $W_bT$. This preserves the transformation and lets us simplify the combined branch operations.

We can also replace a CNOT near the end of $\pre$ with a classically controlled Pauli-$X$ gate when its control is the measured qubit. This is allowed if the operations on the control qubit between the CNOT and measurement are diagonal or anti-diagonal. In the anti-diagonal case, we also relabel the measurement outcome. If these conditions hold only approximately, we check whether the full-circuit loss still remains at most $\epsilon$.

For the configuration in Fig.~\ref{fig:framework}(c), each branch $W_b$ can be absorbed into the subsequent block $B'$, yielding $B'_b=B'\widetilde W_b$. We retain a separate target unitary for each measurement history during subsequent reconstruction.

After synthesizing the branches and applying recursion and these simplifications, we recompute the full-circuit loss and expected CNOT count. We replace the current circuit $C$ with the candidate $C'$ only if the loss is at most $\epsilon$ and the expected CNOT count is lower:
\begin{equation}
 \Delta(\widehat U_{C'},U_T)\leq\epsilon
 \quad\text{and}\quad \cost(C')<\cost(C).\label{eq:accept}
\end{equation}
The pre-measurement search keeps the lowest-error candidate for each budget. Among the original static circuit and the dynamic candidates before and after postprocessing that meet the error threshold, we keep the one with the lowest expected CNOT count.

\subsection{Implementation Optimizations}\label{sec:implementation}

Many candidates share the same circuit context, making repeated full-circuit evaluations costly. We cache a local environment incorporating the fixed circuit context and target unitary, allowing each candidate's full-circuit loss to be evaluated through a smaller matrix calculation.

To build this cache, we write the full circuit as $\widehat U_C=R\widetilde U_{\mathrm{dyn}}L$, where $L$ and $R$ contain the operations before and after the candidate, respectively, and $\widetilde U_{\mathrm{dyn}}$ embeds $\dyn$ into the full circuit. Let $S$ be the set of qubits on which the candidate acts coherently. This set also includes earlier measurement controls. Following QFactor~\cite{qfactor}, we compute and cache the environment matrix
\begin{equation}
 E_S=\tr_{\bar S}(LU_T^\dagger R),\label{eq:environment}
\end{equation}
where $\bar S$ contains the qubits outside $S$ and $\tr_{\bar S}$ is the partial trace over them. By cyclic invariance of the trace, we can then compute the full-circuit overlap as
\begin{equation}
 \tr(U_T^\dagger\widehat U_C)=\tr(\dyn E_S).\label{eq:cachedtrace}
\end{equation}
We use the cached $E_S$ to compute each candidate's global loss and recompute it only when $S$ or the surrounding circuit changes.

\section{Experimental Evaluation}
\label{sec:experiments}

\subsection{Experimental Setup}
\label{sec:experimental_setup}

\subsubsection{Baselines and CAVQS Implementation}
We compare \CAVQS{} with QFAST~\cite{qfast}, SQUANDER~\cite{squander}, and Fast-USYN~\cite{fastusyn}. In the tables, SQUANDER and Fast-USYN are abbreviated as SQ and FU respectively.

To the best of our knowledge, no prior dynamic variational synthesis method directly supports this setting. AC/DC focuses on reducing circuit depth, while its general unitary-generation protocol~\cite{acdc} requires additional ancillas and is excluded from the quantitative comparison. We therefore construct \emph{QFAST-SM} (abbreviated as QF-SM) by appending terminal computational-basis measurements to QFAST's synthesized circuit and applying tket's \texttt{SimplifyMeasured} pass~\cite{tket_simplifymeasured}, which replaces eligible pre-measurement quantum operations with classical post-processing of measurement outcomes. To evaluate Eq.~\eqref{eq:loss}, we restore removed phases and represent classical post-processing coherently. Gate counts refer to the simplified executable circuits.

\CAVQS{} is implemented in Python 3.12.7 using QFAST 2.2.0 for block-based synthesis and QSearch~\cite{qsearch} for gate-level instantiation.\footnote{The source code will be made publicly available upon publication.} Following QFAST, we set the block width to $s=3$ for three-qubit targets and $s=2$ for four- to six-qubit targets. Computational-basis measurements are permitted on all outputs, and no additional ancillas are allowed.

\subsubsection{Benchmark Construction}
We evaluate two classes of targets: random-circuit unitaries and quantum algorithm benchmarks, with all tested methods receiving only the target unitary matrices as inputs.

\emph{Random-circuit targets.}
At each generation step, we apply a randomly parameterized $U3$ gate to a randomly selected qubit, followed with probability 0.8 by a CNOT on a selected qubit pair. The numbers of steps are $4,6,\ldots,16$ for three and four qubits, $4,6,\ldots,12$ for five qubits, and $4,6,8,10$ for six qubits. We generate ten circuits for each qubit-step pair and extract their unitary matrices, yielding 230 targets in total.

\emph{Quantum algorithm benchmarks.}
We use all 29 four- to six-qubit target unitary matrices from the public QFAST benchmark suite~\cite{qfast}. The benchmarks include the quantum Fourier transform (QFT), quantum imaginary-time evolution (QITE), variational quantum eigensolver (VQE), quantum approximate optimization algorithm (QAOA), multiplier, and transverse-field Ising model (TFIM) circuits.

Both the ablation and connectivity experiments are conducted on random-circuit targets with four to six qubits. The ablation experiments use all 160 targets, while the connectivity experiments use 30 targets, comprising ten randomly selected circuits for each qubit count (four, five, and six).

\subsubsection{Evaluation Protocol and Metrics}
All experiments run on an Intel Core i9-13900K workstation with 48\,GB RAM and Windows 11. The accuracy threshold is $\epsilon=10^{-3}$ for the loss in Eq.~\eqref{eq:loss}. Each method or configuration runs independently three times per target, with a one-hour limit per run. Runs that exceed this limit, fail the accuracy test, or encounter a memory error are unsuccessful and are excluded from the gate-count and time averages.

We report the two-qubit gate count $G$, synthesis time $T$ in seconds, and successful/total target counts $S$. For each method, we average successful runs within each target and then average these per-target means. A dash denotes an unsupported case or one with no successful result.

Each CNOT or controlled-Z (CZ) counts as one two-qubit gate. For Fast-USYN, which uses parameterized controlled-$R_z$ (CRZ) gates, we report counts as $x$--$2x$, where $x$ is the native CRZ count and $2x$ is the maximum CNOT count after decomposition. For \CAVQS{}, $G$ is the expected two-qubit gate count in Eq.~\eqref{eq:cost}, evaluated on the uniform-superposition reference input $|+\rangle^{\otimes n}$.

For each comparison, $R$ denotes \CAVQS{}'s relative reduction in mean two-qubit gate count against baseline $A$, computed on their common successful targets as
$
R=
\left(1-
{\sum_{i\in\mathcal I_A}G_{\mathrm{CAVQS},i}} /
{\sum_{i\in\mathcal I_A}G_{A,i}}
\right)\times100\%,
$
where $\mathcal I_A$ contains the targets successfully synthesized by both baseline $A$ and \CAVQS{}, and $G_{A,i}$ and $G_{\mathrm{CAVQS},i}$ are their respective mean gate counts over successful runs on target $i$.

\subsection{Overall Synthesis Results}
\paragraph{Random-circuit targets}
\begin{table}[t]
\centering
\caption{Random-circuit results.}
\label{tab:random}
\footnotesize
\setlength{\tabcolsep}{1.5pt}
\settowidth{\resultintegerwidth}{-00}
\resizebox{\columnwidth}{!}{%
\begin{tabular}{@{}ccccc@{\hspace{3pt}}cccc@{}}
\toprule
\multirow{2}{*}{Method} &
\multicolumn{4}{c}{3 qubits} &
\multicolumn{4}{c}{5 qubits}\\
\cmidrule(lr){2-5}\cmidrule(lr){6-9}
 & \multicolumn{1}{c}{$S$}
 & \multicolumn{1}{c}{$G$}
 & \multicolumn{1}{c}{$T$}
 & \multicolumn{1}{c}{$R$}
 & \multicolumn{1}{c}{$S$}
 & \multicolumn{1}{c}{$G$}
 & \multicolumn{1}{c}{$T$}
 & \multicolumn{1}{c}{$R$}\\
\midrule
QFAST & 70/70 & \resultentry{8}{01}{} & 59.54 & \resultentry{40}{90}{\%}
      & 50/50 & \resultentry{20}{94}{} & 109.54 & \resultentry{34}{71}{\%}\\
QF-SM & 70/70 & \resultentry{8}{01}{} & 59.68 & \resultentry{40}{90}{\%}
      & 50/50 & \resultentry{20}{94}{} & 109.92 & \resultentry{34}{71}{\%}\\
SQ & 70/70 & \resultentry{9}{99}{} & 2.65 & \resultentry{52}{62}{\%}
   & 49/50 & \resultentry{13}{69}{} & 22.12 & \resultentry{2}{05}{\%}\\
FU & --\textsuperscript{a} & \multicolumn{1}{c}{--} & -- & \multicolumn{1}{c}{--}
   & 35/50 & \resultentry{47}{02}{--94.04} & 1973.55 & \resultentry{74}{64}{--87.32\%}\\
\CAVQS{} & 70/70 & \resultentry{4}{73}{} & 236.00 & \multicolumn{1}{c}{N/A}
          & 50/50 & \resultentry{13}{67}{} & 97.12 & \multicolumn{1}{c}{N/A}\\
\midrule
\multirow{2}{*}{Method} &
\multicolumn{4}{c}{4 qubits} &
\multicolumn{4}{c}{6 qubits}\\
\cmidrule(lr){2-5}\cmidrule(lr){6-9}
 & $S$ & \multicolumn{1}{c}{$G$} & $T$ & \multicolumn{1}{c}{$R$}
 & $S$ & \multicolumn{1}{c}{$G$} & $T$ & \multicolumn{1}{c}{$R$}\\
\midrule
QFAST & 70/70 & \resultentry{19}{69}{} & 38.49 & \resultentry{28}{79}{\%}
      & 39/40 & \resultentry{22}{83}{} & 1131.04 & \resultentry{42}{52}{\%}\\
QF-SM & 70/70 & \resultentry{19}{69}{} & 38.83 & \resultentry{28}{79}{\%}
      & 39/40 & \resultentry{22}{83}{} & 1131.54 & \resultentry{42}{52}{\%}\\
SQ & 70/70 & \resultentry{21}{26}{} & 49.13 & \resultentry{34}{06}{\%}
   & 0/40\textsuperscript{b} & \multicolumn{1}{c}{--} & -- & \multicolumn{1}{c}{--}\\
FU & 69/70 & \resultentry{19}{72}{--39.44} & 576.14 & \resultentry{30}{51}{--65.26\%}
   & 0/40\textsuperscript{b} & \multicolumn{1}{c}{--} & -- & \multicolumn{1}{c}{--}\\
\CAVQS{} & 70/70 & \resultentry{14}{02}{} & 38.06 & \multicolumn{1}{c}{N/A}
          & 39/40 & \resultentry{12}{79}{} & 954.83 & \multicolumn{1}{c}{N/A}\\
\bottomrule
\end{tabular}%
}
\par\vspace{2pt}
\begin{minipage}{\columnwidth}
\fontsize{6pt}{7pt}\selectfont\raggedright
\textsuperscript{a}Fast-USYN does not support three-qubit synthesis.
\par
\textsuperscript{b}Both SQUANDER and Fast-USYN timed out on all 40 six-qubit targets under the one-hour time limit.
\par
$G$ averages across each method's own successful targets, whereas $R$ is computed on targets successfully synthesized by both CAVQS and the corresponding baseline. Thus, $R$ cannot generally be calculated from the displayed $G$ values.
\end{minipage}
\end{table}

Table~\ref{tab:random} shows a consistent improvement over QFAST: \CAVQS{} reduces mean two-qubit gate counts by 28.79\%--42.52\% while matching its success counts. QFAST-SM leaves two-qubit gate counts unchanged, although it can still remove a small number of single-qubit gates (not reported in Table~\ref{tab:random}). This is because the QSearch-instantiated circuits expose few measurement-based simplification opportunities directly exploitable by \texttt{SimplifyMeasured}, whereas CAVQS actively creates such opportunities through dynamic block reconstruction. For four- to six-qubit targets, these reductions accompany comparable or shorter synthesis times than QFAST; the additional synthesis cost is concentrated in the three-qubit case.

At five qubits, \CAVQS{}'s gate-count advantage narrows to 2.05\%, while SQUANDER synthesizes circuits substantially faster. The five-qubit random-circuit targets are generally shallower than the four-qubit targets, which may favor SQUANDER's gate-block removal and reoptimization strategy. Fast-USYN produces larger circuits and requires much longer synthesis times at both four and five qubits, with fewer successful targets. At six qubits, \CAVQS{} completes 39 of 40 targets, whereas SQUANDER and Fast-USYN complete none.

\paragraph{Quantum algorithm benchmarks}
Table~\ref{tab:applications} indicates that \CAVQS{} achieves the best overall gate counts (4q/5q/6q: 14.40 / 20.65 / 16.80) while maintaining strong success rates (100\% / 83.3\% / 71.4\%). On the commonly successful targets, \CAVQS{} yields average gate reductions of 27.27\% (4q), 30.70\% (5q), and 36.77\% (6q) versus QFAST/QFAST-SM. Meanwhile, SQUANDER and Fast-USYN fail to scale to 6q (both 0\% success), highlighting \CAVQS{}’s better robustness at higher qubit counts. Beyond the overall results, \CAVQS{} matches $G=17$ on QFT(4), reduces QITE(4) to 29 gates (vs.~33), cuts Multiplier(5) from 34 to 11, and achieves $G=15.50$ on TFIM(6) (vs.~22).

\begin{table}[t]
\centering
\caption{Quantum algorithm benchmark results.}
\label{tab:applications}
\footnotesize
\setlength{\tabcolsep}{2pt}
\resizebox{\columnwidth}{!}{%
\begin{tabular}{@{}cccccccccc@{}}
\toprule
\multirow{2}{*}{Target ($x$, $n$)} &
\multicolumn{2}{c}{QFAST} &
\multicolumn{2}{c}{QF-SM} &
\multicolumn{2}{c}{SQ} &
\multicolumn{2}{c}{FU} &
\multicolumn{1}{c}{\multirow{2}{*}{\CAVQS{}}}\\
\cmidrule(lr){2-3}
\cmidrule(lr){4-5}
\cmidrule(lr){6-7}
\cmidrule(lr){8-9}
 & \multicolumn{1}{c}{$G$}
 & \multicolumn{1}{c}{$R$}
 & \multicolumn{1}{c}{$G$}
 & \multicolumn{1}{c}{$R$}
 & \multicolumn{1}{c}{$G$}
 & \multicolumn{1}{c}{$R$}
 & \multicolumn{1}{c}{$G$}
 & \multicolumn{1}{c}{$R$}
 & \\
\midrule
QFT (1, 4) & 17 & 0.00\%
 & 17 & 0.00\%
 & 17 & 0.00\%
 & 21--42 & 19.05--59.52\% & 17\\
QITE (1, 4) & 33 & 12.12\%
 & 33 & 12.12\%
 & 96 & 69.79\%
 & 31--62 & 6.45--53.23\% & 29\\
QAOA (1, 5) & 47 & 25.53\%
 & 47 & 25.53\%
 & 20 & -75.00\%
 & -- & -- & 35\\
VQE (1, 5) & 40 & 37.50\%
 & 40 & 37.50\%
 & -- & --
 & -- & -- & 25\\
TFIM-5-10 (1, 5) & 20 & 42.50\%
 & 20 & 42.50\%
 & 50 & 77.00\%
 & -- & -- & 11.50\\
Multiplier (1, 5) & 34 & 67.65\%
 & 34 & 67.65\%
 & 20 & 45.00\%
 & 51--102 & 78.43--89.22\% & 11\\
TFIM-6-10 (1, 6) & 22 & 29.55\%
 & 22 & 29.55\%
 & -- & --
 & -- & -- & 15.50\\
\midrule
4q overall (10, 4) & 19.80 & 27.27\%
 & 19.80 & 27.27\%
 & 46.60 & 69.10\%
 & 19.89--39.78 & 36.31--68.16\%
 & 14.40\\
5q overall (12, 5) & 31.64 & 30.70\%
 & 31.64 & 30.70\%
 & 38.50 & 20.65\%
 & 48.00--96.00 & 71.88--85.94\%
 & 20.65\\
6q overall (7, 6) & 17.00 & 36.77\%
 & 17.00 & 36.77\%
 & -- & --
 & -- & --
 & 16.80\\
\bottomrule
\end{tabular}%
}
\par\vspace{2pt}
\begin{minipage}{\columnwidth}
\fontsize{6pt}{7pt}\selectfont\raggedright
In Target ($x$, $n$), $x$ denotes the number of benchmark targets and $n$ the number of qubits.
The upper half lists representative circuits selected from the full test set, while the lower half summarizes results aggregated by qubit count. As in Table~\ref{tab:random}, $R$ is computed on common successful targets for CAVQS and each baseline.
\par
Success rates for 4q, 5q, and 6q, respectively:
QFAST/QF-SM: 100\%, 91.7\%, 28.6\%;
SQ: 100\%, 50\%, 0\%;
FU: 90\%, 16.7\%, 0\%;
CAVQS: 100\%, 83.3\%, 71.4\%.
All 6q SQ and FU runs timed out.
\end{minipage}
\end{table}

\subsection{Ablation Study}

\begin{table}[t]
\centering
\caption{Component ablation results on random-circuit targets.}
\label{tab:ablation}
\fontsize{7.5pt}{9pt}\selectfont
\renewcommand{\arraystretch}{0.96}
\setlength{\tabcolsep}{2pt}
\setlength{\aboverulesep}{0.3ex}
\setlength{\belowrulesep}{0.4ex}
\begin{tabular*}{\columnwidth}{@{\extracolsep{\fill}}cccccc@{}}
\toprule
$n$ & Metric
 & \CAVQS{}-U
 & \CAVQS{}-D
 & \CAVQS{}-SM
 & \CAVQS{}\\
\midrule
\multirow{3}{*}{4}
 & $G$ & 16.63 & 17.00 & 16.63 & 14.02\\
 & $T$ & 27.34 & 47.55 & 27.61 & 38.06\\
 & $R$ & 15.70\% & 17.55\%
 & 15.70\% & N/A\\
\midrule
\multirow{3}{*}{5}
 & $G$ & 16.60 & 17.51 & 16.60 & 13.67\\
 & $T$ & 104.63 & 122.39 & 104.92 & 97.12\\
 & $R$ & 17.64\% & 21.92\%
 & 17.64\% & N/A\\
\midrule
\multirow{3}{*}{6}
 & $G$ & 16.32 & 18.72 & 16.32 & 12.79\\
 & $T$ & 981.27 & 1146.96 & 981.65 & 954.83\\
 & $R$ & 21.63\% & 29.09\%
 & 21.63\% & N/A\\
\bottomrule
\end{tabular*}
\par\vspace{2pt}
\begin{minipage}{\columnwidth}
\fontsize{6.5pt}{7.5pt}\selectfont\raggedright
All variants succeed on 70/70 (4q), 50/50 (5q), and 39/40 (6q) targets, but their unsuccessful 6q targets are not identical.
\end{minipage}
\end{table}

In Table~\ref{tab:ablation}, CAVQS-U retains the static part of CAVQS without dynamic reconstruction, whereas CAVQS-D applies dynamic reconstruction to circuits synthesized by unmodified QFAST. CAVQS-SM appends \texttt{SimplifyMeasured} to CAVQS-U.

Combining the two core steps reduces two-qubit gate counts by at most 21.63\% relative to CAVQS-U and 29.09\% relative to CAVQS-D across the common successful four- to six-qubit targets. Relative to CAVQS-U, CAVQS-SM leaves two-qubit gate counts unchanged. The full pipeline is also faster than both CAVQS-U and CAVQS-D at five and six qubits.

\subsection{Connectivity Extension}
\begin{table}
\centering
\caption{Connectivity-constrained synthesis results.}
\label{tab:connectivity}
\fontsize{7.77pt}{9.2pt}\selectfont
\setlength{\tabcolsep}{2pt}
\begin{tabular*}{\columnwidth}{@{\extracolsep{\fill}}cccccccccc@{}}
\toprule
\multirow{2}{*}{$n$} &
\multirow{2}{*}{Topology} &
\multicolumn{3}{c}{QFAST} &
\multicolumn{3}{c}{SQ} &
\multicolumn{2}{c}{\CAVQS{}}\\
\cmidrule(lr){3-5}
\cmidrule(lr){6-8}
\cmidrule(lr){9-10}
 & & $G$ & $S$ & $R$
     & $G$ & $S$ & $R$
     & $G$ & $S$\\
\midrule
4 & Line & 29.50 & 10/10 & 45.55\%
  & 15.90 & 10/10 & -1.03\%
  & 16.06 & 10/10\\
  & Ring & 20.40 & 10/10 & 21.05\%
  & 14.90 & 10/10 & -8.09\%
  & 16.11 & 10/10\\
  & Star & 20.50 & 10/10 & -1.95\%
  & 12.13 & 8/10 & -34.02\%
  & 20.90 & 10/10\\
\midrule
5 & Line & 37.67 & 9/10 & 35.35\%
  & 20.14 & 7/10 & -14.89\%
  & 28.12 & 10/10\\
  & Ring & 39.00 & 10/10 & 40.77\%
  & 17.29 & 7/10 & -9.09\%
  & 23.10 & 10/10\\
  & Star & 27.90 & 10/10 & 25.28\%
  & 16.75 & 8/10 & -14.52\%
  & 20.85 & 10/10\\
\midrule
6 & Line & 39.00 & 2/10 & 43.59\%
  & -- & 0/10 & --
  & 19.33 & 3/10\\
  & Ring & 33.71 & 7/10 & 37.08\%
  & -- & 0/10 & --
  & 21.31 & 8/10\\
  & Star & 24.00 & 9/10 & 12.79\%
  & -- & 0/10 & --
  & 18.75 & 8/10\\
\bottomrule
\end{tabular*}
\end{table}

Following QFAST~\cite{qfast}, we achieve connectivity-constrained synthesis by restricting candidate block placements according to the target coupling graph. We exclude Fast-USYN from this comparison because its implementation does not yet fully support connectivity-constrained synthesis. Table~\ref{tab:connectivity} shows aggregate two-qubit gate-count reductions over QFAST across all three topologies, with the largest gains on line and ring graphs, exceeding one third. Although SQUANDER achieves the lowest mean gate counts at four and five qubits, it finds no solutions in the six-qubit tests. \CAVQS{} successfully synthesizes all four- and five-qubit targets and produces successful results on every topology at six qubits, with higher success rates than QFAST on line and ring graphs.

\section{Conclusion}
We presented \CAVQS{}, an ancilla-free synthesis framework combining QR-based unitary redistribution with dynamic reconstruction using CNOT-budgeted search and SVD-based branch construction. Experiments show lower expected two-qubit gate counts on random-circuit and quantum algorithm benchmarks. Ablations demonstrate the complementary contributions of the two steps, while connectivity experiments confirm gate-count savings under connectivity constraints.

\section*{Content Generated by AI}
Large language models assisted only with language polishing. The authors produced the scientific content, reviewed all assisted text, and take full responsibility for the paper.

\clearpage
\bibliographystyle{IEEEtran}
\bibliography{references}

\end{document}